\documentclass[superscriptaddress, twocolumn, showpacs, amsmath, amssymb, aps, pra, floatfix, 10pt]{revtex4-2}
\usepackage[american]{babel}
\usepackage[T1]{fontenc}
\usepackage{hyperref}
\usepackage{graphicx}
\usepackage{booktabs}
\usepackage{colortbl}
\usepackage{multirow}
\usepackage{physics}
\usepackage{tikz}
\usetikzlibrary{decorations.pathmorphing}
\usetikzlibrary{decorations.pathreplacing}
\hypersetup{colorlinks=true, urlcolor=blue, citecolor=blue, filecolor=blue, linkcolor=blue}

\begin{document}

\title{Wichmann-Kroll Correction in  Muonic Atoms and Hydrogen-Like Electronic Ions: a Comparative Study of Two Methods}

\date{\today}
\author{Zoia~A.~Mandrykina}
\affiliation{Max Planck Institute for Nuclear Physics,  Saupfercheckweg 1, 69117 Heidelberg, Germany}
\author{Zewen~Sun}
\affiliation{Max Planck Institute for Nuclear Physics,  Saupfercheckweg 1, 69117 Heidelberg, Germany}
\author{Natalia~S.~Oreshkina}
\affiliation{Max Planck Institute for Nuclear Physics,  Saupfercheckweg 1, 69117 Heidelberg, Germany}

\begin{abstract}
Wichmann-Kroll corrections are calculated in both hydrogen-like electronic ions and muonic systems ($Z = \{36$--$92\}$) using two independent methods. The Gaussian finite basis set approach, enhanced with dual basis construction, analytical large-distance corrections, and $B$-spline representations, provides computational efficiency. The Green function method, based on semi-analytical construction from Dirac solutions with Fermi nuclear charge distributions, offers higher systematic accuracy and freedom from basis-dependent artifacts. Results are consistent with the literature values, providing reliable reference data for precision spectroscopy of exotic atoms.
\end{abstract}

\maketitle

\section{Introduction}
\label{sec:Introduction}
Precision spectroscopy of atomic systems provides one of the most stringent tests of quantum electrodynamics (QED) in strong Coulomb fields. For highly charged ions and exotic atoms, the comparison between theoretical predictions and experimental measurements has reached a level of accuracy where higher-order QED corrections become essential~\cite{Indelicato2019, Morgner2023}.

Vacuum polarization (VP) represents one of the largest QED corrections 
to atomic energy levels, especially in muonic systems. The leading-order 
VP effect, known as the Uehling correction~\cite{PhysRev.48.55}, has a known analytical form and can now be calculated with high precision for both point-like and 
finite-sized nuclei~\cite{Huang1976Calculation, Klarsfeld1997Analytical}. 
In contrast, the next-order contribution, 
the Wichmann-Kroll (WK) correction~\cite{Wichmann1956Vacuum}, which corresponds to the $(Z\alpha)^3$ 
and higher terms in the VP expansion, requires the evaluation of the 
electron propagator in the presence of the nuclear field. This makes 
the WK correction significantly more challenging to compute and necessitates demanding
numerical approaches.

The WK correction becomes increasingly important for high-$Z$ systems, where it can contribute several electronvolts to the binding energies of inner-shell electrons~\cite{Persson1993Accurate, Mohr1998, Soff1988}. For muonic atoms, where the muon orbits much closer to the nucleus due to its larger mass, VP effects are drastically enhanced, and WK contribution reaches hundreds of electronvolts~\cite{Rinker1975Vacuum}. Recent interest in muonic atoms spectroscopy, driven by the proton radius puzzle~\cite{Pohl2010, Antognini2013} and ongoing measurements of the nuclear charge radii~\cite{Krauth2021}, has renewed the demand for accurate WK calculations across a wide range of nuclear charges.

There are two principal approaches {when} calculating the WK correction. The first employs finite basis set (FBS) methods to construct the VP charge density from discrete representations of the Dirac spectrum. This approach, particularly when implemented with Gaussian basis functions, offers computational efficiency and has been recently demonstrated to achieve competitive accuracy~\cite{Ivanov2024Vacuum, Salman2023Calculating}. The second approach uses the Dirac Green function (GF) constructed from regular and irregular solutions of the radial Dirac equation~\cite{Soff1988, Mohr1998, Yerokhin2011Nuclear}. This method provides higher intrinsic accuracy but requires careful numerical treatment and is in general computationally more demanding.

In this work, we present a comprehensive comparison of the Gaussian FBS and semi-analytical method to construct the Green functions in electronic and muonic systems across a range of nuclear charges $Z = \{36 - 92\}$ for a few first low-lying levels. 
We compare the two  approaches, and discuss their accuracy and efficiency under different scenarios. 
In addition, we provide refined numerical methods for the calculation of the WK-induced charge density 
adopting Gaussian basis set, improving the accuracy and efficiency of the calculations. 
Finally, we systematically present the Green function method, and provided a comparison between the results from the two methods. 
This allows us to evaluate the 
trade-off between computational efficiency and accuracy, and to provide practical guidance for choosing the 
appropriate method depending on the system under study.

The paper is organized as follows. In Sec.~\ref{sec:basicformalism}, we present the general theoretical framework for WK calculations. Section~\ref{sec:FBS_method} describes the Gaussian basis set implementation and details the FBS method along with our numerical improvements. Section~\ref{sec:Green_construction} presents the semi-analytical approach to constructing the Green function. Our results are discussed in Sec.~\ref{sec:Results}, and conclusions are given in Sec.~\ref{sec:conсlusion}.

Throughout the paper, we employ natural units with $\hbar = c = 1$ and with Heaviside charge units $e^2 = 4\pi\alpha$. The unit of mass is chosen according to the system under consideration: $m_\mu = 1$ for muonic atoms and $m_e = 1$ for electronic ions.

\section{Basic formalism}
\label{sec:basicformalism}
\begin{figure}[t!]
\centering
\includegraphics[width=1\linewidth]{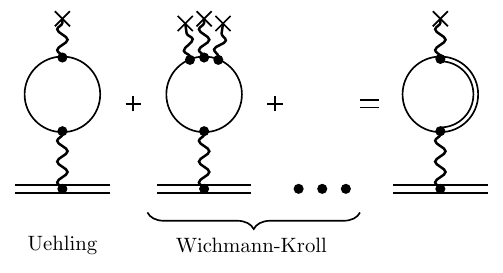}
\caption{Feynman diagram for vacuum polarization correction decomposition. The double lines indicate the bound particle and virtual electron-positron state in the field of the nucleus, the wavy lines represent virtual photon emission.}
\label{fig:wk_feynman}
\end{figure}
The energy shift due to vacuum polarization for a bound state with 
principal quantum number $n$ and relativistic angular quantum number $\kappa$ is given by
\begin{equation}
\label{eq:E_corre}
    \Delta E_{n\kappa} = 
    \int d
    \boldsymbol{r}
    \psi^\dagger_{n\kappa m}(\boldsymbol{r})
    V_{\rm VP}(\boldsymbol{r})
    \psi_{n\kappa m}(\boldsymbol{r}),
\end{equation}
where $\psi_{n\kappa m}$ is the wave function of the bound particle represented as a bispinor form~\cite{Akhiezer1965}:
\begin{equation}
    \psi_{n\kappa m}(\boldsymbol{r})
    =
    \dfrac{1}{r}
    \begin{pmatrix}
        G_{n\kappa}(r)\Omega_{\kappa m}(\boldsymbol{\hat{r}}) \\
        i F_{n\kappa}(r)\Omega_{-\kappa m}(\boldsymbol{\hat{r}})
    \end{pmatrix}.
\end{equation}
Here, $G_{n\kappa}(r)$ and $F_{n\kappa}(r)$ are the large and small 
radial components, and $\Omega_{\pm\kappa m}$ are the spherical spinors with the projection of the total angular number $m$.
The vacuum polarization potential $V_{\rm VP}$ is related to the 
induced charge density $\rho_{\rm VP}$:
\begin{equation}
    V_{\rm VP}(\boldsymbol{r})
    =
    - \alpha \int
    d\boldsymbol{x}
    \frac{\rho_{\rm VP}(\boldsymbol{x})}{|\boldsymbol{r}-\boldsymbol{x}|}.
\end{equation}
The vacuum polarization charge density can be expressed as 
\begin{equation}
    \rho_{\rm VP}(\boldsymbol{x})
    =
    \dfrac{1}{2\pi i}
    \int_{C_F} 
    d \omega ~ 
    {\rm Tr \mathcal{G}}(\omega,\boldsymbol{x}, \boldsymbol{x}),
\end{equation}
where $\mathcal{G}(\omega, \boldsymbol{x}, \boldsymbol{x})$ is the Dirac 
Green function in the presence of the nuclear potential, and the integration over $\omega$ runs along the Feynman contour $C_F$. An alternative representation via the spectral decomposition of the Green function reads:
\begin{align}
    \rho_\mathrm{VP}(\boldsymbol{x}) =\frac{1}{2} &\Big[ \sum_{E_{n}>0} \mathrm{Tr} \left( \psi_{n}^{\dagger}(\boldsymbol{x}) \psi_{n}(\boldsymbol{x}) \right) \notag \\
    &- \sum_{E_{n}<0} \mathrm{Tr} \left( \psi_{n}^{\dagger}(\boldsymbol{x}) \psi_{n}(\boldsymbol{x}) \right) \Big], 
\end{align}
where the sums run over positive- and negative-energy eigenstates, respectively.

The Green function can be expanded in powers of the binding potential (see Fig.~
\ref{fig:wk_feynman}), which leads to the corresponding expansion of the charge density in powers of~$(Z\alpha)$:
\begin{multline}
\label{eq:rho_VP}
    \rho_{\rm VP}(\boldsymbol{x})
    =
    \rho_{\rm VP}^{(1)}(\boldsymbol{x})
    +
    \rho_{\rm VP}^{(3)}(\boldsymbol{x})
    +
    \rho_{\rm VP}^{(5)}(\boldsymbol{x})
    +
    \dotso
    \\
    =
    \rho_{\rm VP}^{(1)}(\boldsymbol{x})
    +
    \rho_{\rm VP}^{(3+)}(\boldsymbol{x})
    =
    \rho_{\rm Ue}(\boldsymbol{x})
    +
    \rho_{\rm WK}(\boldsymbol{x}),    
\end{multline}
where $\rho_{\rm VP}^{(i)}$ denotes the term of order $(Z\alpha)^i$. 
Only odd powers appear due to Furry's theorem~\cite{Peskin1995}. The first-order term 
corresponds to the Uehling correction, while the higher-order terms 
$(i \geqslant 3)$ constitute the Wichmann-Kroll contribution.

It should be noted that $\rho_\mathrm{VP}^{(1)}(\boldsymbol{x})$ formally contains ultraviolet divergences, which is not carried by $\rho_\mathrm{VP}^{(3+)}(\boldsymbol{x})$. 
Therefore, charge renormalization~\cite{Peskin1995} with respect to $\rho_\mathrm{VP}^{(1)}(\boldsymbol{x})$ offers well-defined $\rho_\mathrm{VP}^{(3+)}(\boldsymbol{x})$.

The two approaches, discussed in details below, correspond to different ways of constructing the Dirac Green function. 
The first method constructs the Green function using wavefunctions directly solved from FBS. 
This method, due to its simplicity, is considered a robust and numerically efficient but inaccurate method. 
The second method constructs the Green function by combining the regular solutions at the origin and at infinity, as discussed in Refs.~\cite{Soff1988, Yerokhin2011Nuclear, Zaytsev2024qed}. This  method provides higher accuracy but is computationally more demanding. 
The details of each approach are presented in Secs.~\ref{sec:FBS_method} and \ref{sec:Green_construction}, respectively.

\section{Finite-basis-set Method}
\label{sec:FBS_method}
{
Recently, Salman and Saue~\cite{Salman2023Calculating} demonstrated the feasibility of using the FBS method with Gaussian basis to achieve relatively accurate WK calculations. 
Later, Ivanov \textit{et al.}~\cite{Ivanov2024Vacuum} continued this method and compared Gaussian and $B$-spline basis, showing Gaussian significantly outperforms $B$-spline basis. 
In this paper, we introduce numerical techniques to further improve Gaussian FBS method. 
We apply a dual basis construction~(DBC) method to tame the imperfect oscillation behaviors of WK charge density at large distance. 
We also provide an analytic way to correct the charge density at this region, resulting in more accurate outcomes. 
In addition, using $B$-spline representation of the charge density, we reduced the calculation time of WK correction by orders of magnitude. }

\subsection{Gaussian basis}
A Gaussian basis set is commonly used in atomic physics and chemistry, and has recently been demonstrated as a suitable basis for QED calculations~\cite{Salman2023Calculating, Ivanov2024Vacuum, Ferenc2025Gaussian}. 
It is defined as
\begin{equation}
    \pi_{i}^{\pm}(r) = \mathcal{N} r^{d_{\pm}} \exp \left(-\zeta_{i} r^{2} \right), 
\end{equation}
where $\mathcal{N}$ is the normalization factor. 
The $r^{d_{\pm}}$ and exponential terms characterize the asymptotic behavior of the wavefunction at $r\rightarrow0$ and $+\infty$, respectively. 
For realistic nuclear models, the charge distribution is roughly a constant at $r\rightarrow0$, therefore the asymptotic wavefunction is similar to the free-particle one, such that
\begin{equation}
    d_{\pm} = \left|\kappa \pm \frac{1}{2} \right| + \frac{1}{2}. 
\end{equation}
The $\zeta_i$ coefficients must follow certain conditions for the basis functions to be sufficiently linearly independent, and in this work, we use
\begin{equation}
\label{eq:zeta}
    \zeta_{i} = \zeta_{1} \left(\zeta_{N} / \zeta_{1} \right)^{\frac{i-1}{N-1}}, 
\end{equation}
where $N$ is the number of basis functions. 
The coefficients $\zeta_1$ and $\zeta_N$ are determined based on
\begin{align}
\label{eq:zeta1}
    \zeta_1 = 1 / \left( 2 \, r_1^2 \right),& \ \zeta_N = 1 / \left( 2 \, r_N^2 \right), \\
\label{eq:zeta2}
    r_1 = r_\mathrm{rms} / 20,& \ r_N = r_\mathrm{B}/7, 
\end{align}
where $r_\mathrm{rms}$ is the root mean square (rms) nuclear charge radius, and $r_\mathrm{B}$ is the Bohr radius. 
This choice of $\{ \zeta_i \}$ is empirically determined, showing a good performance with both small and large $Z$.

The radial wavefunctions are expressed as
\begin{align}
    G(r) &= \sum_{i=1}^N c_i \, \pi_{i}^{+}(r) \\
    F(r) &= \sum_{i=1}^N c_{i+N} \, \pi_{i}^{-}(r), 
\end{align}
such that solving the radial Dirac equation is now converted to solving the $2N$ coefficients $\{ c_i \}$. 
{The coefficients are calculated using the Rayleigh–Ritz variational method, e.g.~explained in Ref.~\cite{Grant2007Relativistic, Salman2023Calculating}. }
It should be noted that some properties of Gaussian basis can significantly simplify this problem (see Appendix~\ref{apdx:numerical_methods}).

\subsection{WK-induced charge density}
Vacuum polarization can be treated as the energy correction caused by the VP-induced charge density $\rho_\mathrm{VP}^{(i)} (r)$ introduced in Eq.~(\ref{eq:rho_VP}). 
For FBS method, $\rho_\mathrm{VP}^{(i)}(r)$ is constructed directly using wavefunctions solved from the Gaussian basis. 
Studies have shown the Gaussian basis is more appropriate than the $B$-spline basis for such calculations~\cite{Ivanov2024Vacuum}. 
The method can be summarized as the following:
\begin{align}
\label{eq:rho}
    \rho_\mathrm{VP}(r) 
    &= \sum_{\kappa = \pm 1}^{ \pm \infty} \rho_{\kappa}(r) \\
    \rho_{\kappa}(r) 
    &= |\kappa| \sum_{n} \operatorname{sgn}\left(E_{n, \kappa}\right) \rho_{n, \kappa}(r) \\
\label{eq:rho_nk}
    \rho_{n, \kappa}(r) 
    &= \psi_{n, \kappa}^{\dagger} \psi_{n, \kappa} = \frac{1}{4\pi r^2} \left( F_{n, \kappa}^{2} + G_{n, \kappa}^{2} \right), 
\end{align}
where the factors $\abs{\kappa}$ and $1/4\pi$ are related to the angular wavefunctions, and $F_{n, \kappa}$ and $G_{n, \kappa}$ are numerically solved radial wavefunctions.

Charge conjugation symmetry, also denoted as $\mathcal{C}$ symmetry, must be ensured in vacuum polarization calculation as described in Ref.~\cite{Salman2020Charge, Salman2023Calculating}. 
However, numerical solutions obtained via FBS method do not automatically satisfy this $\mathcal{C}$ symmetry.
Implementing the dual kinetic balance (DKB) approach~\cite{Shabaev2004DKB} can ensure the $\mathcal{C}$ symmetry. 
Alternatively, a more straightforward method,  used also in this work, is adopting 
\begin{equation}
\label{eq:rho_kc}
    \rho_{ \kappa, \mathcal{C} }(r, Z) = \frac{1}{2} \left[ \rho_\kappa (r, Z) - \rho_\kappa (r, -Z) \right], 
\end{equation}
where $\rho_{ \kappa, \mathcal{C} }$ is the VP-induced charge density satisfying $\mathcal{C}$ symmetry.

The WK correction is the higher order term of VP, as shown in Eq.~(\ref{eq:rho_VP}), therefore
\begin{equation}
    \rho_\mathrm{VP}^{(3+)} = \rho_\mathrm{VP} - \rho_\mathrm{VP}^{(1)}. 
\end{equation}
Since $\rho_\mathrm{VP}^{(1)}$ corresponds to the linear term, in numerical calculation, this equation becomes
\begin{equation}
\label{eq:rho_k3}
    \rho_{\kappa}^{(3+)}(r, Z) = \rho_{\kappa}(r, Z) - \lim _{\delta \rightarrow 0} \frac{Z}{\delta} \rho_{\kappa}(r, \delta). 
\end{equation}
Overall, $\rho_\mathrm{VP}^{(3+)}(r)$ can be calculated by combining Eqs.~\eqref{eq:rho}-\eqref{eq:rho_kc}, and \eqref{eq:rho_k3}. 

\begin{figure}[h]
    \centering
    \includegraphics[width=0.46\textwidth]{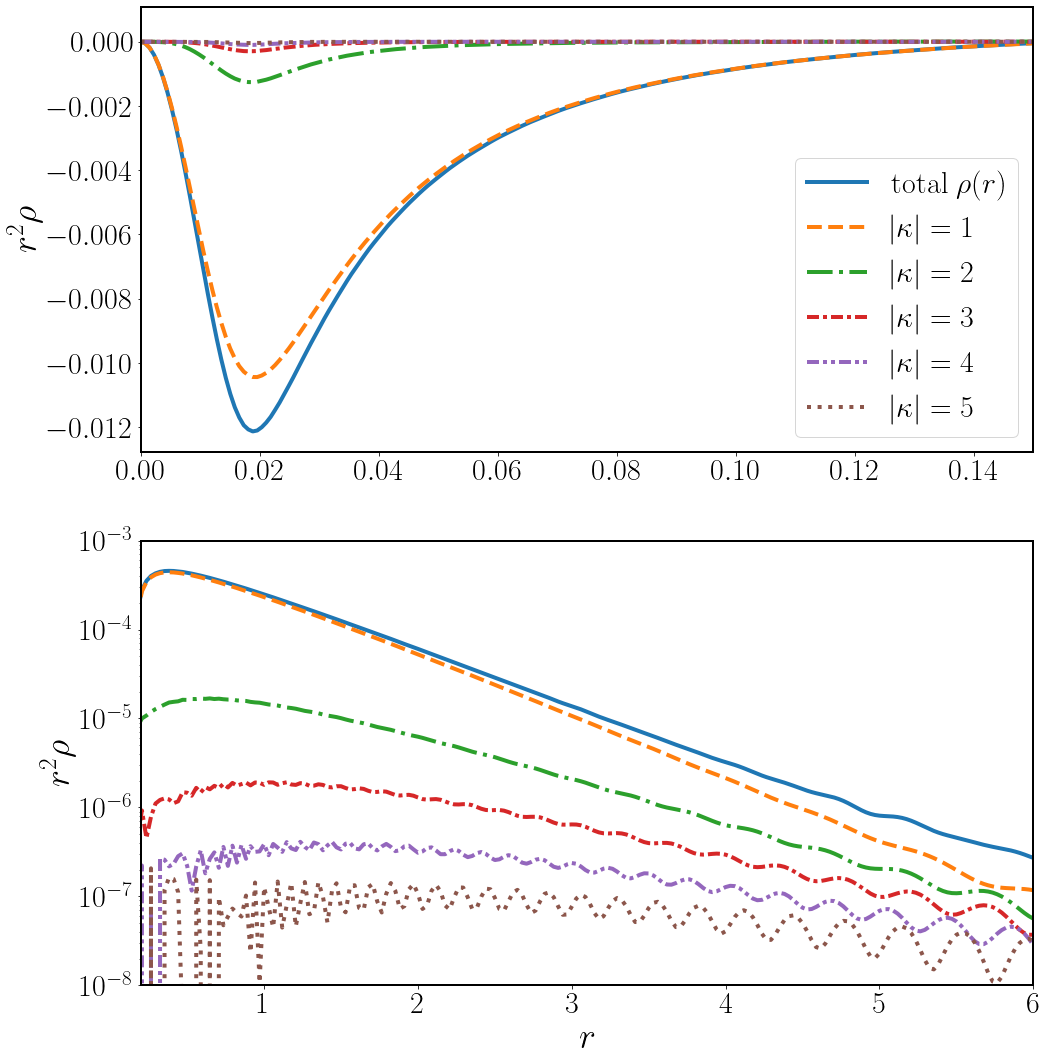}
    \caption{WK charge density $\rho_\mathrm{VP}^{(3+)}(r)$ of $_{92}$U with different $\abs{\kappa}$ contributions at small (upper) and large (lower) distance, in natural (electronic) unit. }
    \label{fig:rho}
\end{figure}

\begin{figure}[h]
    \centering
    \includegraphics[width=0.46\textwidth]{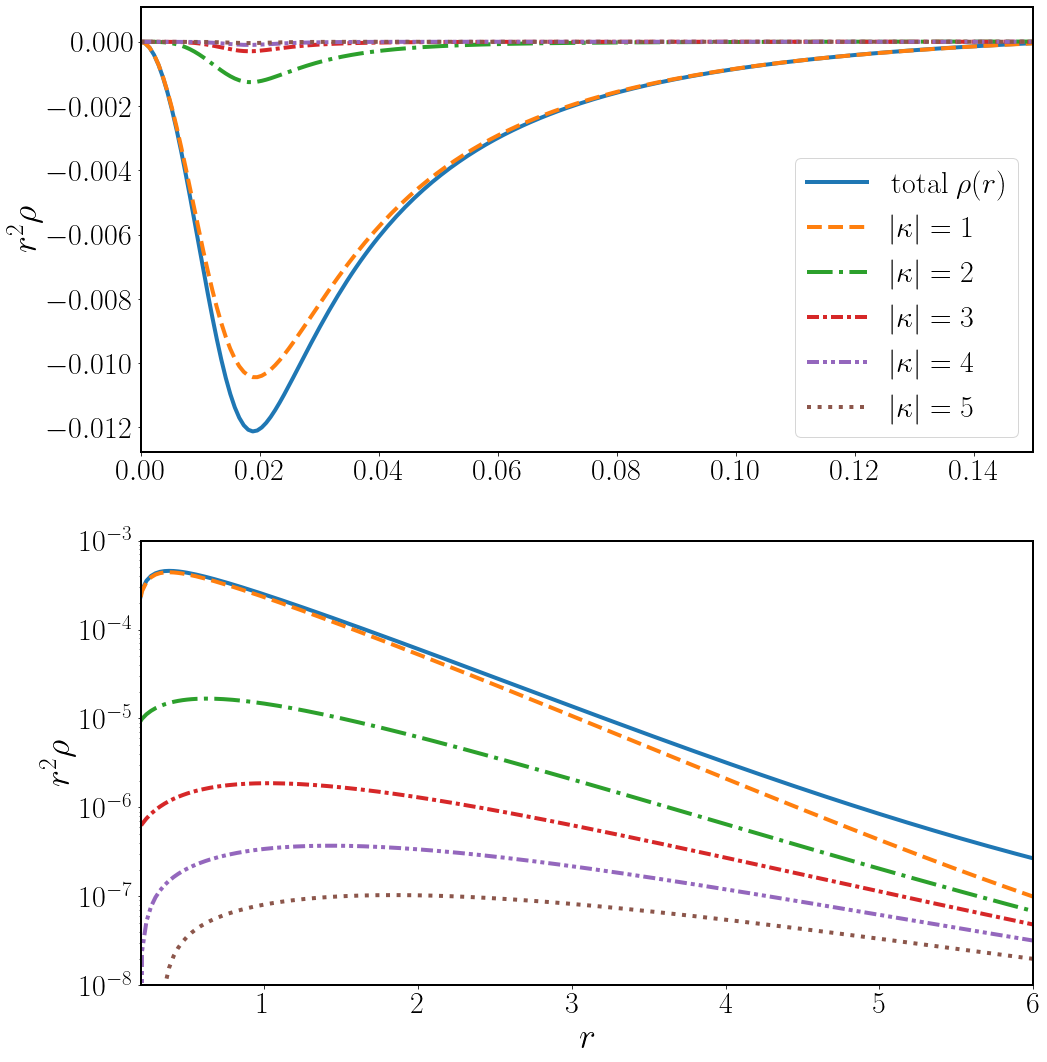}
    \caption{WK charge density $\rho_\mathrm{VP}^{(3+)}(r)$ of $_{92}$U calculated using dual basis construction~(DBC) method. }
    \label{fig:rho_DBC}
\end{figure}

The 
result of $\rho_\mathrm{VP}^{(3+)}(r)$
for $_{92}$U calculated from a spherical nuclear model with rms radius of 5.860~fm is presented in Fig.~\ref{fig:rho}. 
It shows that $\rho_\mathrm{VP}^{(3+)}(r)$
has small but incorrect oscillatory behavior, particularly for large $\abs{\kappa}$. 
This oscillation can be mitigated by increasing the number of basis functions, but cannot be completely eliminated. 
In this paper, we introduce the DBC method to address it. 
Since this oscillation is basis dependent, i.e.~changing the positions and densities of basis functions will alter the oscillation, one can construct a second basis with ``opposite oscillation phase'' to cancel the oscillation. 
To do this, we choose $2N$ numbers of $\{ \zeta_i \}$ coefficients according to Eqs.~\eqref{eq:zeta}-\eqref{eq:zeta2}, and then separate them into two sets, even labeled $\{ \zeta_{2i} \}$ and odd labeled $\{ \zeta_{2i+1} \}$. 
This creates two sets of basis, each having $N$ number of functions. 
We calculate $\rho_\kappa(r)$ twice using dual basis and take the average result, which exhibits a smooth plot, as shown in Fig.~\ref{fig:rho_DBC}. 
{Using this DBC method, we can achieve a faster convergence of the resulting WK energy corrections as a function of  $N$.}

\subsection{WK correction and convergence}
As shown in Eq.~(\ref{eq:rho_VP}), Eq.~(\ref{eq:E_corre}) calculates the WK energy correction when considering only the higher-order charge density term $3+$. The corresponding WK potential is given by
\begin{align}
    V_\mathrm{WK}(r) 
    = - 4\pi \alpha \bigg[& \frac{1}{r} \int_0^{r} {r'}^2 \rho_\mathrm{VP}^{(3+)} \left( r' \right) \mathrm{d}r' \notag \\
    &+ \int_r^\infty r' \rho_\mathrm{VP}^{(3+)} \left( r' \right) \mathrm{d}r' \bigg]. 
\end{align}
In order to efficiently calculate $V_\mathrm{WK}(r)$, we first numerically create a $B$-spline {approximation} of $\rho_\mathrm{VP}^{(3+)}(r)$ on a sufficiently dense grid of $r$. 
This method does not lead to a sacrifice in accuracy, since the grid is sufficiently dense, and at the same time significantly improves the calculation speed by multiple orders of magnitude. 
This method converts $\rho_\mathrm{VP}^{(3+)}(r)$ into spline knots and coefficients, from which the potential energy is calculated. 
The potential energy can be used to calculate WK energy shift for both electronic atomic system and exotic atoms with muons, and the results are shown in the following Sec.~\ref{sec:Results}.

The WK potential energy calculated from FBS method is slightly inaccurate at large distance, since the asymptotic behavior of Gaussian basis function at $r \rightarrow \infty$ is $\exp( -\zeta r^2 )$, whereas for $\rho_\mathrm{VP}^{(3+)}(r)$ should be $\exp( -\eta r )$, where $\zeta$ and $\eta$ are some coefficients. 
Meanwhile, it is not feasible to construct sufficiently dense basis functions at large distance to reduce this incorrectness.  
Here, we propose an efficient method to tackle this problem. 
Since the potential energy for a point and finite-sized nuclei should converge at large $r$, one can use the WK potential of point-like nucleus to correct the finite-sized nucleus calculations. 
The WK potential of point-like nuclei is studied in Ref.~\cite{Fainshtein1991Vacuum}, and we smoothly connect their expression with our calculated $V_\mathrm{WK}(r) $ at large $r$. 
The difference in results with and without this analytic correction method are presented in Appendix~\ref{apdx:numerical_methods}. 

Regarding numerical convergence, the results converge as the basis 
approaches completeness. Ideally, the basis becomes complete when 
$\zeta_N \rightarrow 0$, $(\zeta_1/\zeta_N)^{1/(N-1)} \rightarrow 1^+$, 
and $N \rightarrow \infty$. We investigate these three criteria numerically.

The practical choice of $\{\zeta_i\}$ given in Eqs.~\eqref{eq:zeta}--\eqref{eq:zeta2} 
ensures that $\zeta_N$ is sufficiently small for the results to reach 
convergence. A previous study~\cite{Ferenc2025Gaussian} suggests 
$(\zeta_1/\zeta_N)^{1/(N-1)} \simeq 1.4$--$1.5$; our choice yields smaller 
values of approximately $1.1$--$1.2$ at $N = 160$, indicating a denser 
basis coverage.

We therefore focus on the third criterion and quantify the convergence 
with respect to $N$. Increasing $N$ corresponds to increasing the spatial 
density of basis functions, which can lead to numerical issues when basis 
functions become nearly linearly dependent at finite floating-point precision. 
To overcome this problem, we employ arbitrary-precision arithmetic using 
the \texttt{mpmath} Python library~\cite{mpmath}. 
The calculations are performed with $N = 130$, $140$, $150$, and $160$ 
to confirm the convergence. 

\section{Green function method}
\label{sec:Green_construction}
\subsection{Construction of the Dirac Green functions}
In contrast to the FBS method described in Sec.~\ref{sec:FBS_method}, 
the Green function approach constructs $\mathcal{G}(\omega, \boldsymbol{r}, 
\boldsymbol{r'})$ directly from the solutions of the radial Dirac equation. 
This method requires combining solutions that are regular at the origin 
($r \to 0$) with those regular at infinity ($r \to \infty$), matched at 
an intermediate point. While computationally more demanding, this approach 
avoids the spurious oscillations and basis completeness issues discussed 
in Sec.~\ref{sec:FBS_method}. Here, the electron is treated as the loop particle.

The WK potential in this formalism is given by~\cite{Zaytsev2024qed}:
\begin{align}
    V_{\rm WK}(\boldsymbol{r})
    & = -\alpha \int d\boldsymbol{r'}\dfrac{\rho^{(3+)}_{\rm VP}(\boldsymbol{r'})}{|\boldsymbol{r} - \boldsymbol{r'}|} =
    \dfrac{\alpha}{2\pi i}
    \int d\boldsymbol{r'}\dfrac{1}{|\boldsymbol{r} - \boldsymbol{r'}|} \notag \\
    & \times \int_{C_{F}}d\omega {\rm Tr}\biggl[
    \int d\boldsymbol{x} \mathcal{G}^{(0)}(\omega,\boldsymbol{r'},\boldsymbol{x})
    V(x) \notag \\
    & \times \left(
    \mathcal{G}(\omega, \boldsymbol{x}, \boldsymbol{r'}) - \mathcal{G}^{(0)}(\omega,\boldsymbol{x},\boldsymbol{r'})
    \right)
    \biggr].
\label{greenwk}
\end{align}
In Eq.~(\ref{greenwk}),
$\mathcal{G}^{(0)}(\omega,\boldsymbol{r_1},\boldsymbol{r_2})$ 
represents the free-electron Dirac Green function, 
$\mathcal{G}(\omega, \boldsymbol{r_1}, \boldsymbol{r_2})$ denotes the Dirac Green function in the presence of an external potential, and $V(x)$ is the binding potential.
\\
The Dirac Green function for a free particle can be constructed as:
\begin{multline}
    \mathcal{G}^{(0)}(E, \boldsymbol{r_1}, \boldsymbol{r_2})
    =
    -\sum_{\kappa m}
    \Bigg[
    \Phi^\infty_{E\kappa m}(\boldsymbol{r_1})\Phi^{0\dagger}_{E\kappa m}(\boldsymbol{r_2})\theta(r_1-r_2) \\
    +
    \Phi^{0}_{E\kappa m}(\boldsymbol{r_1})\Phi^{\infty\dagger}_{E\kappa m}(\boldsymbol{r_2})\theta(r_2-r_1)   
    \Bigg],
\label{greenfun}
\end{multline}
where the functions 
$\Phi^\infty_{E\kappa m}$ and $\Phi^{0}_{E\kappa m}$
are expressed as:
\begin{equation}
    \Phi_{E\kappa m}(\boldsymbol{r})
    =
    \dfrac{1}{r}
    \begin{pmatrix}
        G_{E\kappa}(r)\Omega_{\kappa m}(\boldsymbol{\hat{r}}) \\
        i F_{E\kappa}(r)\Omega_{-\kappa m}(\boldsymbol{\hat{r}}).
    \end{pmatrix}
\end{equation}
Here, $\kappa$ is the Dirac quantum number, $m$ is the magnetic quantum number, $\Omega_{\pm\kappa m}(\boldsymbol{\hat{r}})$ are the spherical spinors.
The big and small components, $ G_{E\kappa}(r)$ and $F_{E\kappa}(r)$, satisfy the radial Dirac equation for a free particle:
\begin{equation}
    \begin{pmatrix}
    \dfrac{d}{dr} + \dfrac{\kappa}{r} & m_ec^2+E \\
    m_ec^2-E & \dfrac{d}{dr} - \dfrac{\kappa}{r}
    \end{pmatrix}
        \begin{pmatrix}
        G_{E\kappa}(r)\\
        F_{E\kappa}(r)
    \end{pmatrix}
    = 0.
\end{equation}
\\
\\
The solutions that are regular at the origin can be expressed in terms of spherical Bessel functions:
\begin{equation}
    \begin{pmatrix}
        G^{0}_{E\kappa}(r)\\
        F^{0}_{E\kappa}(r)
    \end{pmatrix}
    =
    r
    \begin{pmatrix}
        j_{|\kappa+1/2|-1/2}(ipr) \\
        i\dfrac{p}{|m_ec^2+E|}\dfrac{\kappa}{|\kappa|}j_{|\kappa-1/2|-1/2}(ipr)
    \end{pmatrix}
\end{equation}

Similarly, the solutions that are regular at infinity are given in terms of spherical Hankel functions of the first kind:
\begin{equation}
    \begin{pmatrix}
        G^{\infty}_{E\kappa}(r)\\
        F^{\infty}_{E\kappa}(r)
    \end{pmatrix}
    =
    r
    \begin{pmatrix}
        h^{(1)}_{|\kappa+1/2|-1/2}(ipr) \\
        i\dfrac{p}{|m_ec^2+E|}\dfrac{\kappa}{|\kappa|}h^{(1)}_{|\kappa-1/2|-1/2}(ipr)
    \end{pmatrix},
\end{equation}
where $p=\sqrt{(E/c)^2-(m_e c)^2}$ and $m_e$ are the momentum and the mass of the electron, respectively.
For the external-potential Dirac Green function, we employ the same construction as in the free-particle case, but with $\Phi$ representing the solution of the Dirac equation in the presence of the external potential $V(r)$:
\begin{equation}
    \begin{pmatrix}
    \dfrac{d}{dr} + \dfrac{\kappa}{r} & m_ec^2+E-V(r) \\
    m_ec^2-E+V(r) & \dfrac{d}{dr} - \dfrac{\kappa}{r}
    \end{pmatrix}
        \begin{pmatrix}
        G_{E\kappa}(r)\\
        F_{E\kappa}(r)
    \end{pmatrix}
    = 0.
\end{equation}
$\Phi^{0}_{E\kappa m}(\boldsymbol{r})$ is obtained by the propagation from $r=0$, while the function $\Phi^{\infty}_{E\kappa m}(\boldsymbol{r})$ was constructed by propagation from the $r>R$ to $r=0$. In this case $R$ is the point where the nuclear potential turns into the pure Coulomb one. And for the case with $r>R$ we can employ the following analytical solution:
\begin{align}
    &\begin{pmatrix}
        G^{\infty}_{E\kappa}(r)\\
        F^{\infty}_{E\kappa}(r)
    \end{pmatrix}
    =
    \dfrac{1}{\sqrt{2pr}} \times \\
    &\begin{pmatrix}
        \left(\kappa + \dfrac{\nu~m_e c^2}{E}\right)
        W_{-}(2pr)+W_{+}(2pr)\\
        \dfrac{p}{m_e c^2+E}\left[ \left(
        \kappa + \dfrac{\nu~m_e c^2}{E}
        \right)
        W_{-}(2pr)-W_{+}(2pr)
        \right]
    \end{pmatrix}. \notag
\end{align}
Here, $p$ is the momentum of the electron as defined previously, $\nu = \alpha ZE/p$ is the Sommerfeld parameter (where $\alpha$ is the fine-structure constant and $Z$ is the nuclear charge number), and $\gamma = \sqrt{\kappa^2-(\alpha Z)^2}$ is the relativistic angular momentum parameter. The functions $W_{\pm}(z)$ represent the Whittaker functions $W_{\nu\pm 1/2,\gamma}(z)$, respectively.

\begin{table*}[t]
\centering
\begin{tabular}{ccccccccccccc}
\hline \hline
 & rms (fm) & bound particle & $\kappa = |1|$ & $\kappa = |2|$ & $\kappa = |3|$ & $\kappa = |4|$ & $\kappa = |5|$ & $\kappa = |6|$ & $\kappa = |7|$ & $\kappa = |8|$ & $\kappa = |9|$ & $\kappa = |10|$\\
 \hline
$_{92}$U & 5.8571 & muon & 635.5797 & 50.0414 & 8.8026 & 2.3802 & 0.8469 & 0.3695 & 0.1922 & 0.1183 & 0.0853 & 0.0704\\
 & & electron & 4.4689 & 0.3934  & 0.0814  & 0.0248 & 0.0096 & 0.0044 & 0.0023 & 0.0013 & 0.0009  & 0.0007\\
\hline \hline
\end{tabular}
\caption{Partial wave contributions to the WK correction (in eV) for the $1s_{1/2}$ state of muonic and electronic uranium ($Z = 92$), calculated using the GF method.}
\label{tab:kappa_convergence}
\end{table*}

\begin{table*}[t]
\centering
\newcolumntype{G}{!{\color{gray}\vrule}} 
\begin{tabular}{cc G cccc G cccc G c}
\arrayrulecolor{black}
\hline \hline
 & & \multicolumn{4}{c G}{Electronic systems} & \multicolumn{4}{c G}{Muonic systems} & \\
  & rms(fm) & $1s_{1/2}$ & $2s_{1/2}$ & $2p_{1/2}$ & $2p_{3/2}$ & $1s_{1/2}$ & $2s_{1/2}$ & $2p_{1/2}$ & $2p_{3/2}$ & Method \\
\hline
\arrayrulecolor{gray}
 $_{36}$Kr & 4.230 & 0.01550 & 0.00200 & 0.00006 & 0.00002 & 36.63 & 11.31 & 14.75 & 14.35 & FBS\\
           &  & 0.01550 & 0.00200 & 0.00006 & 0.00002 &  &  &  &  & $b$\\
           &  & $0.01553(1)$ & $0.001997(3)$ & $0.000072(3)$ & $0.0000345(6)$ &  $36.82(5)$ & $11.34(2)$ & $14.75(2)$ & $14.35(2)$ & GF\\
\hline
 $_{54}$Xe & 4.7859 &   &  &  &  &  $137.3[9]$ & $53.0[2]$ & $74.3[1]$ & $71.5[0] $ & $a$ \\
           & 4.7964 & $0.1697(1)$ & $0.02302(1)$ & $0.001652(2)$ & $0.0004347(9)$ & $137.0(2)$ & $52.95(7)$ & $74.18(9)$ & $71.39(9)$ & GF\\
            & 4.826  & 0.1695 & 0.0230 & 0.0016 & 0.0004 & 135.5 & 52.58 & 73.99 & 71.25 & FBS\\
           &  & 0.1695 & 0.0230 & 0.0016 & 0.0004 &   &  &  &  & $a$ \\

\hline 
 $_{70}$Yb & 5.273 & 0.8283(3) & 0.1198 & 0.0153 & 0.0028 & 302.7 & 136.9 & 197.0 & 189.6 & FBS\\
           &  & 0.8283 & 0.1198 & 0.0153 & 0.0028 &   &  &  &  & $b$ \\
           & 5.3215 & $0.8284(5)$ & $0.12001(6)$ & $0.01550(3)$ & $0.002965(8)$ & $302.3(5)$ & $136.9(2)$ & $196.2(3)$ & $188.8(2)$ & GF\\
\hline 
 $_{82}$Pb & 5.5012  & 2.2901[3] & 0.3534[0] & 0.0660[0] & 0.0093[-1] &  &  &  &  & $a$ \\
           &  & $2.291(1)$ & $0.3539(2)$ & $0.0665(1)$ & $0.00968(5)$ & $500.52(8)$ & $247.2(3)$ & $353.0(5)$ & $340.1(4)$ & GF \\
           & 5.505  & 2.2901(8)  & 0.3534(1) & 0.0661 & 0.0094 & $496.8(1)$ & 245.8 & 352.4 & 339.7 & FBS\\           
           & 5.51  &  &  &  &  & 492 & 244 & 348 & 335 & $c$ \\
\hline
 $_{92}$U   & 5.8571 & $4.988(2)$  & $0.8221(4)$ & $0.206(2)$ & $0.0232(1)$ & $697.7(1)$ & $370.9(5)$ & $523.4(7)$ & $506.5(7)$ & GF \\
           &   & 4.9870[10] & 0.8215[2] & 0.2056[0] & 0.0225[0] &  698.3[47] & 371.2[22] & 523.8[11] & 506.9[7] & $a$ \\
            & 5.860  & $4.9859(11)$ & $0.8212(2)$ & $0.2058(1)$ & 0.0226  & 692.8 & 368.7 & 522.2 & 505.8 & FBS\\
           &   & 4.9863 & 0.8214 & 0.2057 & 0.0226 &   &  &  &  & $b$ \\         
           & 5.72 & &  &  & &  691 & 367 & 517 & 500  & $c$ \\
\arrayrulecolor{black}
\hline \hline
\end{tabular}
\vspace{0.3em}
    
\raggedright
\small
FBS: Sec.~\ref{sec:FBS_method}, sphere model. \\
GF: Sec.~\ref{sec:Green_construction}, Fermi model, $|\kappa|_{\mathrm{max}} = 5$ for muonic systems and $|\kappa|_{\mathrm{max}} = 10$ for electronic systems.\\
$a$: Ref.~\cite{Jonas2025All}, Fermi model, $|\kappa|_\mathrm{max} = 12$. Parenthetical values (in [ ]): subtractive correction for sphere model.\\
$b$: Ref.~\cite{Persson1993Accurate}, sphere model, $|\kappa|_\mathrm{max} \leqslant 5$.\\
$c$: Ref.~\cite{Rinker1975Vacuum}, Fermi model, $|\kappa|_\mathrm{max} = 2$.
\caption{WK correction in eV for $1s_{1/2}$, $2s_{1/2}$, $2p_{1/2}$ and $2p_{3/2}$ states of hydrogen-like ions with $Z = 36, 54, 70, 82, 92$. Left panel: electronic systems; right panel: muonic systems.}
\label{tab:tab_all}
\end{table*}

\subsection{Numerical implementation and partial-wave convergence}

The numerical calculations are implemented in \textsc{Fortran90}. The radial Dirac equation is solved using the method developed by Salvat \textit{et al.}~\cite{Salvat1995}. We employ an integration mesh with exponential grid mapping, which provides optimal resolution near the nucleus where the WK charge density varies rapidly. Numerical errors from integration and spline interpolation are controlled at the $10^{-14}$ level; the dominant source of uncertainty is therefore the truncation of the partial wave expansion at finite~$|\kappa|_{\mathrm{max}}$.

The GF method supports arbitrary nuclear charge distributions. For the present work, we adopt a Fermi model, which provides a realistic description of the nuclear charge density for heavy elements, with parameters fitted to experimental rms radii (Table~\ref{tab:tab_all}).

The Table~\ref{tab:kappa_convergence} presents the partial wave contributions to the WK correction for the $1s_{1/2}$ state of muonic atom and electronic ion. In the case of a muon as a bound particle, the dominant contribution arises from $|\kappa| = 1$, which accounts for 90\% and more of the total correction. Higher partial waves contribute progressively less. 
The rapid convergence justifies truncating the partial-wave expansion at $|\kappa|_{\rm max} = 5$ for most calculations, with an estimated truncation error below 0.2\%. 

For electronic systems, the convergence pattern is qualitatively similar, with the $|\kappa| = 1$ term dominating again. However, the energy corrections are smaller by two orders of magnitude compared to muonic systems. Furthermore, as shown in Table~\ref{tab:tab_all}, the ratio of the muonic to electronic energy correction increases to three to four orders of magnitude for the other orbitals.
At the same time, achieving comparable relative precision requires including more partial waves: we use $|\kappa|_{\mathrm{max}} = 10$ for electronic systems compared to $|\kappa|_{\mathrm{max}} = 5$ for muonic systems.

To estimate the truncation uncertainty, we employed three successive steps: (i) evaluating the relative magnitude of the last included term, (ii) verifying the monotonic decay with increasing $|\kappa|$, and (iii) extrapolating the contribution from the cut-off tail. The extrapolation exploits the observed asymptotic behavior: the analysis of our data reveals consistent power-law decay $C(\kappa) \propto \kappa^{-n}$ with exponents $n \approx 3$--$4$ across all states. We fit the functional form $C(\kappa) = A \kappa^{-n}$ to the final 3--4 calculated terms and evaluate the tail contribution from $|\kappa| > |\kappa|_{\mathrm{max}}$ analytically. This approach provides stable and physically motivated extrapolation, in contrast to polynomial methods which can exhibit spurious oscillations. The resulting truncation uncertainties are indicated in parentheses in the Table~\ref{tab:tab_all}.

\section{Results and discussion}
\label{sec:Results}

We present Wichmann-Kroll   corrections for both muonic atoms and electronic hydrogen-like ions calculated using two independent methods: the finite basis set approach and the Green function construction method. Results are provided for nuclear charges $Z = 36$, $54$, $70$, $82$, and $92$ and states $1s_{1/2}$, $2s_{1/2}$, $2p_{1/2}$, and $2p_{3/2}$, as summarized in Table~\ref{tab:tab_all}.
As expected, the WK corrections for muonic systems exceed those for electronic systems by two to four orders of magnitude, reflecting the greater spatial overlap of the muon wave function with the nuclear charge distribution due to the muon proximity to the nucleus.
Additionally, the results are  sensitive to the choice of nuclear model. For muonic $s$-states, where the wave function significantly penetrates the nuclear region, employing a sphere or a Fermi model can lead to a difference in the energy correction up to several eV. For electronic systems, these effects are considerably smaller. Small discrepancies between our results and literature values~\cite{Jonas2025All, Persson1993Accurate, Rinker1975Vacuum} can be attributed partly to differences in nuclear radii and model assumptions, as well as to the truncation of the partial-wave expansion at different $|\kappa|_{\mathrm{max}}$ values.

In Table~\ref{tab:tab_all}, uncertainties are indicated in parentheses. For the FBS method, these are estimated from the difference between calculations with $N = 150$ and $N = 160$ basis functions. For the GF method, they represent the truncation uncertainty from the $\kappa$-summation, evaluated using the power-law extrapolation procedure.

Our FBS and GF results are in good agreement with the recent calculations of Ref.~\cite{Jonas2025All}. However, a direct quantitative comparison is complicated by differences in the nuclear charge distributions employed: Ref.~\cite{Jonas2025All} uses both sphere and Fermi models with slightly different rms radii, whereas our FBS calculations employ a sphere model and our GF calculations use a Fermi distribution. These differences can account for variations of a few tenths of an eV in electronic systems and several eV in muonic systems, particularly for $s$-states where the wave function penetrates the nuclear region.

\section{Conclusion}
\label{sec:conсlusion}
We have presented Wichmann-Kroll corrections for muonic atoms and electronic hydrogen-like ions ($Z = \{36$--$92\}$) calculated using two independent methods. The agreement between the finite basis set and Green function approaches is typically at the 0.1\% level for electronic systems. For muonic systems, differences of up to 1\% are observed, which can be attributed to the different nuclear models employed (sphere for FBS, Fermi for GF) and slightly different rms radii. 
Systematic uncertainty quantification yields 0.1--0.2\% precision for $s$-states and dominant contributions, with somewhat larger relative uncertainties for $p$-states in light electronic systems where the absolute corrections are very small.
Our results, spanning six orders of magnitude from sub-meV to hundreds of eV, provide reliable reference data for precision spectroscopy of exotic atoms and can serve as benchmarks for future calculations.

{\bf Acknowledgments}. 
Z.A.~M. and Z.~S. contributed equally to this work: Z.~S. performed calculations for muonic and electronic atoms by the FBS method (Sec.~\ref{sec:FBS_method}), while Z.A.~M. carried out calculations by the Green function method (Sec.~\ref{sec:Green_construction}). All authors participated in the analysis and manuscript preparation.
Z.~S. thanks the discussion with R. Benazzouk, D. Ferenc, V. K. Ivanov,  T. Saue, and appreciates the data comparison with J. Sommerfeldt. 
Z.A.~M. thanks V.A.~Zaytsev for providing the initial code framework for Wichmann-Kroll corrections in electronic systems, which was further developed and adapted for this work.

This article comprises parts of the PhD thesis work of Z.A.~M. and Z.~S. to be submitted to Heidelberg University.
N.S.~O. thanks the DFG (German Research Foundation) – Project-ID 273811115 – SFB 1225 ISOQUANT for funding.

\bibliography{refs}


\appendix


\section{Numerical methods}
\label{apdx:numerical_methods}

Integrals with Gaussian basis functions, denoted as $\mathcal{G}(d, \zeta)$, can be expressed in terms of gamma functions $\Gamma(x)$, which read
\begin{align}
\label{eq:G_int}
    \mathcal{G}(d, \zeta) 
    &\equiv \int_0^\infty r^d \exp(-\zeta r^{2}) \, \mathrm{d}r \notag \\
    &= \frac{1}{2} \, \zeta^{-(d+1)/2} \, \Gamma\left( \frac{d+1}{2} \right). 
\end{align}
Integrating over a finite interval can be expressed similarly using incomplete Gamma functions, which can significantly simplify the numerical calculations.

Using the variational method, the radial wavefunctions are expanded in terms of basis functions as
\begin{equation}
    \ket{\pi} \equiv \binom{G(r)}{F(r)} = \sum_{i=1}^{N}\binom{c_i \, \pi^+_i(r)}{c_{i+N} \, \pi^-_i(r)}. 
\end{equation}
In this manner, the Dirac equation is converted to a generalized eigenvalue problem
\begin{equation}
    H_r \ket{\pi} = E \, C \ket{\pi}, 
\end{equation}
where $E$ is the energy eigenvalue of radial Hamiltonian $H_r$ and matrix $C$ are defined and evaluated in the following equations.

The matrix $C$ is block diagonalized with two $N \times N$ matrices
\begin{equation}
    C = 
    \begin{pmatrix}
    C_{11} & 0 \\
    0 & C_{22}
    \end{pmatrix}. 
\end{equation}
The matrix elements are
\begin{align}
    \bra{\pi_i^+} C_{11} \ket{\pi_j^+} &= \mathcal{C}_{ij}^{++} (0) \\
    \bra{\pi_i^+} C_{22} \ket{\pi_j^-} &= \mathcal{C}_{ij}^{--} (0), 
\end{align}
where we have used short-hand notation of 
\begin{equation}
    \mathcal{C}_{ij}^{+-} (x) \equiv \mathcal{G} \left(d^+ + d^- + x, \zeta_i + \zeta_j \right), 
\end{equation}
and $\mathcal{C}_{ij}^{++}(x)$ and $\mathcal{C}_{ij}^{--}(x)$ are defined similarly.

The radial Hamiltonian $H_r$ is 
\begin{equation}
    H_r = 
    \begin{pmatrix}
    H_{11} & H_{12} \\
    H_{21} & H_{22}
    \end{pmatrix}
    = 
    \begin{pmatrix}
    V(r) + m & \frac{\kappa}{r} - \frac{\mathrm{d}}{\mathrm{d} r} \\
    \frac{\kappa}{r}+\frac{\mathrm{d}}{\mathrm{d} r} & V(r) - m 
    \end{pmatrix}, 
\end{equation}
with
\begin{align}
    \bra{\pi_i^+} H_{11} \ket{\pi_j^+} 
    &= V_{ij}^{++} + m \, \mathcal{C}_{ij}^{++} (0) \\
    \bra{\pi_i^+} H_{12} \ket{\pi_j^-} 
    &= \kappa \, \mathcal{C}_{ij}^{+-} (-1) \notag \\
    &- \left[ d^- \, \mathcal{C}_{ij}^{+-} (-1) - 2\zeta_j \, \mathcal{C}_{ij}^{+-} (+1) \right] \\
    \bra{\pi_i^+} H_{21} \ket{\pi_j^-} 
    &= \kappa \, \mathcal{C}_{ij}^{+-} (-1) \notag \\
    &+ \left[ d^+ \, \mathcal{C}_{ij}^{+-} (-1) - 2\zeta_j \, \mathcal{C}_{ij}^{+-} (+1) \right] \\
    \bra{\pi_i^-} H_{22} \ket{\pi_j^-} 
    &= V_{ij}^{--} - m \, \mathcal{C}_{ij}^{--} (0). 
\end{align}
Here, $V_{ij}^{++}$ and $V_{ij}^{--}$ are integrals with nuclear potential energy $V(r)$, defined as 
\begin{equation}
    V_{ij}^{++} = \bra{\pi_i^+} V \ket{\pi_j^+}. 
\end{equation}
For point, shell, sphere, and Gaussian nuclear models, this integral can be calculated using Eq.~\ref{eq:G_int} or similar methods. 
However, for other nuclear models, direct numeral integrations or more sophisticated methods are needed.

After the construction of the Hamiltonian, the radial wavefunctions $F(r)$ and $G(r)$ can be expressed in terms of the coefficients $\{ c_i \}$, and the charge density can be calculated using them. 
The charge density must satisfy the following normalization condition
\begin{equation}
    \int_0^\infty \rho_{\kappa}(r) r^2 \, \mathrm{d}r = 0,
\end{equation}
which can be used for validity check.

\end{document}